\documentclass[manuscript]{aastex62}

\usepackage{epsfig}
\usepackage{color}

\def\apj{{ApJ}}
\def\mnras{{MNRAS}}
\def\aap{{A\&A}}
\def\apjl{{ApJ}}
\def\apjs{{ApJS}}

\def\pasa{{PASA}}

\def\bain{{Bulletin Astronomical Institute of the Netherlands}}
\def\pasj{{Publications of the ASJ}}
\def\jcap{{Journal of Cosmology and Astroparticle Physics}}

\def\msol{M$_{\sun}$}
\def\kms{$\frac{\rm km}{\rm s}$}

\shorttitle{Impact of the Smith Cloud}
\shortauthors{Alig et al.}

\begin{document}

\title[Simulating the impact of the Smith Cloud]{Simulating the impact of the Smith Cloud}

\correspondingauthor{C. Alig}

\author{C. Alig}
\affiliation{Universit\"ats-Sternwarte M\"unchen, Scheinerstr.1, D-81679 M\"unchen, Germany}
\affiliation{Max-Planck-Institut f\"ur extraterrestrische Physik, Postfach 1312, Giessenbachstr., D-85741 Garching, Germany}

\author{S. Hammer}
\affiliation{Universit\"ats-Sternwarte M\"unchen, Scheinerstr.1, D-81679 M\"unchen, Germany}

\author{N. Borodatchenkova}
\affiliation{Universit\"ats-Sternwarte M\"unchen, Scheinerstr.1, D-81679 M\"unchen, Germany}

\author{C. L. Dobbs}
\affiliation{School of Physics and Astronomy, University of Exeter, Stocker Road, Exeter EX4 4QL, UK}

\author{A. Burkert}
\affiliation{Universit\"ats-Sternwarte M\"unchen, Scheinerstr.1, D-81679 M\"unchen, Germany}
\affiliation{Max-Planck-Institut f\"ur extraterrestrische Physik, Postfach 1312, Giessenbachstr., D-85741 Garching, Germany}
\affiliation{Max-Planck-Fellow}

\begin{abstract}

We investigate the future evolution of the Smith Cloud by performing hydrodynamical simulations of the cloud impact onto the gaseous Milky Way Galactic disk. We assume a local origin for the cloud and thus do not include a dark matter component to stabilize it. Our main focus is the cloud's influence on the local and global star formation rate (SFR) of the Galaxy and whether or not it leads to an observable event in the far future. Our model assumes two extremes for the mass of the Smith Cloud, an upper mass limit of 10$^7$\ \msol\ and a lower mass limit of 10$^6$\ \msol\, compared to the observational value of a few 10$^6$\ \msol. In addition, we also make the conservative assumption that
the entirety of the cloud mass of the extended Smith Cloud is concentrated within the tip of the cloud. We find that the impact of the low-mass cloud produces no noticeable change in neither the global SFR nor the local SFR at the cloud impact site within the galactic disk. For the high-mass cloud
we find a short-term (roughly 5 Myr) increase of the global SFR of up to 1\ \msol\ yr$^{-1}$, which nearly doubles the normal Milky Way SFR. This highly localized starburst should be observable.

\end{abstract}


\section{Introduction}
\label{sec:intro}

The Smith Cloud \citep{1963BAN....17..203S}, first named in \cite{1998MNRAS.299..611B}, is a well-known high-velocity cloud (HVC) inside of the Milky Way gaseous halo.
Observations \citep{2008ApJ...672..298W, 2008ApJ...679L..21L, 2009ApJ...703.1832H, 2016ApJ...816L..11F} show that this HVC has an extent of around 3$\times$1 kpc and a total mass of a few 10$^6$ \msol. The cloud's tip is currently positioned at a distance of 7.6 kpc from the Galactic Center with an offset of 2.9 kpc below the Galactic Plane. At its current velocity of around 73 ($\pm$ 26) km s$^{-1}$ toward the Galactic Plane, and a total velocity of roughly 300 km s$^{-1}$, the cloud tip will hit the Galactic Plane in about 27 Myr, assuming a ballistic orbit. It is important to note that the motion toward the galactic plane is an assumption based purely on the cloud morphology.\\

The origin of the Smith Cloud is a matter of debate and ongoing research. The two main models assume either an extragalactic origin or an origin from within the Milky Way itself. Extragalactic origin models include the remnant of a dwarf galaxy or accreting intergalactic gas. In the dwarf galaxy model the Smith Cloud is stabilized by a dark matter halo, as it should have passed through the Milky Way Galactic disk on its current orbit once already \citep{2009ApJ...707.1642N, 2014MNRAS.442.2883N}.
In the work of \cite{2016JCAP...11..021L} the authors argued that the Smith Cloud is an excellent target for the indirect detection of dark matter with radio data due to its vicinity, magnetic field strength, and the amount of dark matter that it should contain. Recently, the future evolution of the Smith Cloud, assuming stabilization by a dark matter halo, has been simulated by \cite{2018MNRAS.473.5514T}. In this Letter the authors find that the cloud is able to pass through the Galactic Disk again in the future, following its current orbit.\\

The second type of model assumes that the cloud originates from the Milky Way System itself. One possible source could be ejection from the Galactic Disk. Ejection mechanisms include a jet-like event \citep{2004PASJ...56..633S} or a galactic fountain \citep{2016arXiv160906309M}. However, it should be noted that this would require
an enormous amount of energy considering the cloud's velocity, mass, and distance from the Galactic Disk. Another possible local origin
could be a condensation from the large reservoirs of cold gas detected within the halos of Milky-Way-like galaxies \citep{2018ApJ...864..132B}.
In the observational paper of \cite{2016ApJ...816L..11F} the authors argued that the Smith Cloud metallicity suggests a disk origin rather than an extragalactic origin. However, in \cite{2017ApJ...837...82H} the authors conducted simulations of the Smith Cloud, resolving the mixing of cloud gas with halo gas, and found that the metallicity of the Smith Cloud may not be a true reflection of its original metallicity. Both the aforementioned work and \cite{2018MNRAS.473.5514T} show that the metallicity of the Smith Cloud does not constrain its origin. In this Letter we present simulations of the future evolution of the Smith Cloud, specifically the impact of the cloud onto the Galactic Disk using a local origin scenario. The main question that we want to answer regards the effect of the impact on the local and global star formation rate (SFR) of the Milky Way.\\

\section{Simulation Model}
\label{sec:mod}

To perform simulations we simplify the cloud model by assuming that the entire mass of the Smith Cloud is concentrated within a sphere that is the size of the cloud tip, which has a diameter of 1 kpc. This overestimates somewhat the mass that will impact onto the galactic plane at once, as some of the mass of the original cloud is distributed within the 3 kpc tail that will fall onto the galactic plane at a later time. We also assume a total mass of 10$^7$ \msol\, which is the high end of the mass estimates for the Smith Cloud \citep{2016IAUS..315....9L}. These two assumptions should increase the likelihood of the cloud impact influencing the strength of the local or even global SFR peak. On the other hand, a negative result for this simplified model should be a limit for any effect on the Milky Way SFR peak by the original, much more extended cloud. In addition, we also test
a model with a cloud mass of 10$^6$\ \msol. We assume in this work that the cloud has a local origin (without establishing a specific case), therefore we ignore any dark matter component.\\

A simple estimate, achieved by comparing the mean cloud density to the mean interstellar medium (ISM) density in the Galactic Disk, already shows that the impact of the cloud onto the ISM will not be 
strong. With a cloud mass of $10^{6}$ \msol\ we get a mean density of around
$1.29 \times 10^{-25}$ g cm$^{-3}$ and a surface density of $1.27$ \msol\ pc$^{-2}$ compared to the ISM density of $10^{-24}$ g cm$^{-3}$ and a mean Milky Way surface density of 10 \msol\ pc$^{-2}$. The mean density 
and surface density of our $10^7$ \msol\ cloud are comparable to the ISM density and Milky Way
surface density, which might be a hint that the cloud originates from Milky Way
in the event that the actual mass of the cloud is higher than currently observed.
A more detailed analysis of the Smith Cloud's survivability can be found in \cite{2009IAUS..254..241B}.\\

Our galaxy simulation is based on the work of \cite{2006AAS...20721704D}, \cite{2008MNRAS.385.1893D} and \cite{2011MNRAS.417.1318D}.
We have implemented this model into the Smoothed Particle Hydrodynamics Code GADGET3 \citep{2005MNRAS.364.1105S}. The galactic potentials that we employ contain a disk component and a time-dependent spiral density pattern. For the spiral potential we only take the part provided by the stellar background as it contains considerably more mass than the gas component. 
The disk is modeled by the logarithmic potential \citep{btbook}:

\begin{equation}
\Phi_L = \frac{1}{2}v_0^2 {\rm ln} \left(R_c^2 + R^2 + \frac{z^2}{q_{\Phi}^2}\right) 
\end{equation}

where $R_c$ is the core halo radius, $v_0$ is the velocity of the flat rotation curve, $q_{\Phi}$ is the axis ratio of the equipotential surfaces, and $R = \sqrt{x^2 + y^2}$ is the radial variable in the equatorial plane.
The values for the parameters that we adapt from \cite{2006AAS...20721704D} are $R_c$ = 1kpc, $q_{\Phi}$ = 0.9 and $v_0 = 215$ \kms.
For the spiral density pattern we use the time-dependent potential \citep{2002ApJS..142..261C}:

\begin{equation}
\Phi_S = - 4\pi G H\rho_0 {\rm exp} \left(-\frac{r - r_0}{R_s}\right)\sum_{n=1}^{3}\left(\frac{C_n}{K_n D_n}\right){\rm cos}(n\gamma)
\end{equation}

where the components dependent upon radius are given by $K_n = \frac{nN}{r {\rm sin}(\alpha)}$, $D_n = \frac{1 + K_nH + 0.3(K_n H)^2}{1 + 0.3 K_n H}$ and the time-plus-radius-dependent component is given by 
$\gamma = N\left[\theta - \Omega_p t - \frac{{\rm ln}(r/r_0)}{{\rm tan}(\alpha)} \right]$.
The variable $N$ determines the number of arms, $\alpha$ the pitch angle, $R_s$ the radial scale length of the
drop-off in density amplitude of the arms, $\rho_0$ the mid-plane arm density at fiducial radius $r_0$ and finally $H$ the scale height of the stellar arm perturbation. The values for the constants are given by $C(1) = 8/3\pi$,
$C(2) = 1/2$, and $C(3) = 8/15\pi$. The values for all of the parameters that we again adapt from \cite{2006AAS...20721704D} are
$r_0 = 8$ kpc, $R_s = 7$ kpc, $H = 0.18$ kpc, $\alpha = 15^\circ$, $\Omega_p = 2\times10^{-8}$ rad yr$^{-1}$, and $\rho_0 = 1$ cm$^{-3}$.
All of the galactic potential parameters are tuned to be Milky-Way-like.
We employ an adiabatic equation of state with an adiabatic exponent of $\gamma = 5/3$ and a
mean molecular weight of $\mu = 1.27$, which corresponds to a standard mixture of roughly $71.1\%$ atomic hydrogen, $27.41\%$ helium, and $1.49\%$ metals. The cooling function is a simple parametrization, adapted from \cite{2007ApJ...657..870V}.\\

\begin{figure}
\begin{center}
\includegraphics[width=17cm]{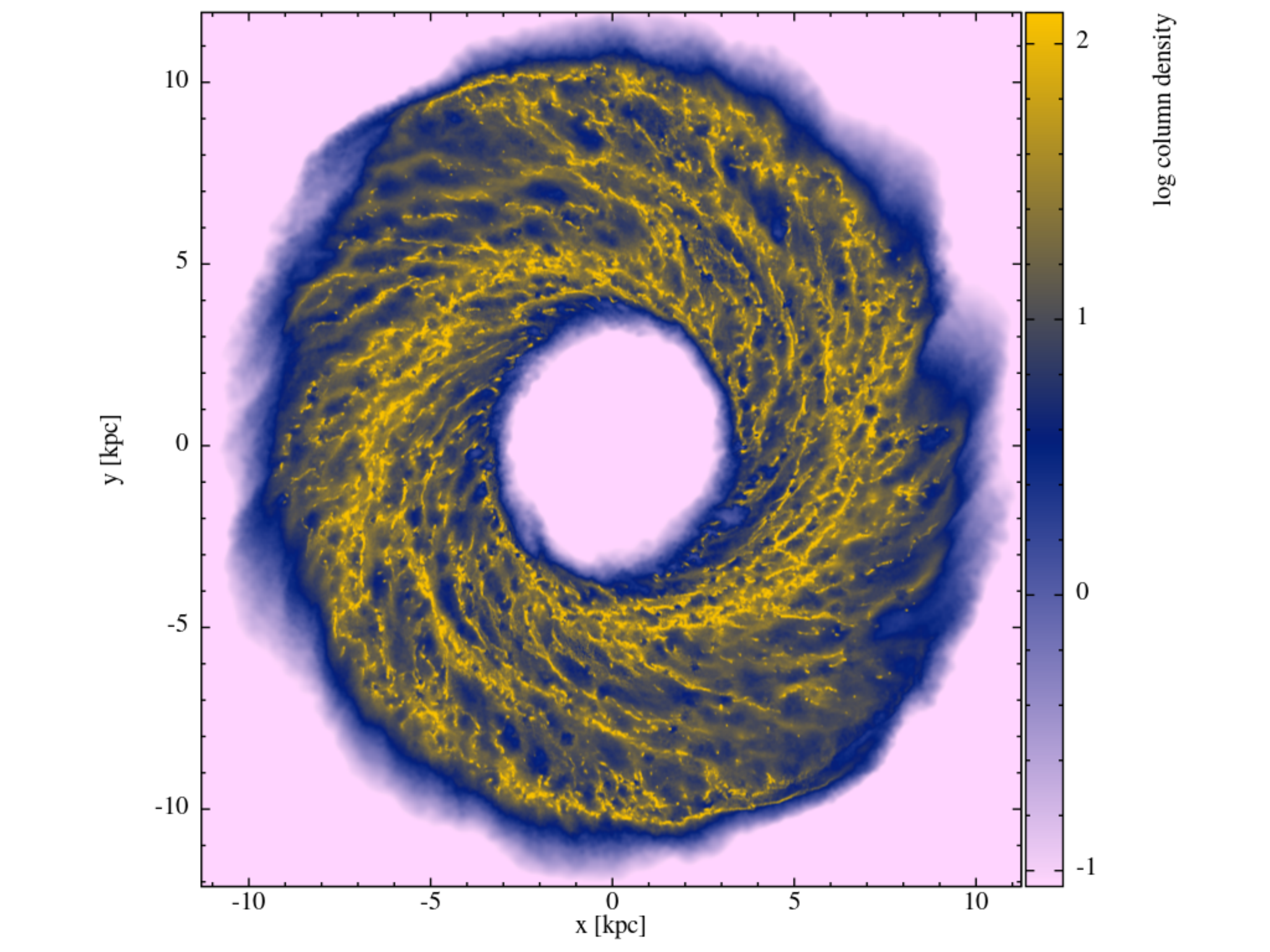}
\end{center}
\caption{
Surface density in \msol\ pc$^{-2}$ for a Milky-Way-like galaxy simulated with our model. The snapshot is taken at 500 Myr from the initial conditions and shows the equilibrium state at which a roughly constant SFR of 1.5 \msol\ yr$^{-1}$ has been established, which is comparable to the Milky Way SFR. This state represents the starting point for our Smith Cloud impact simulations.
}
\label{fig:galinit}
\end{figure}

\begin{figure}
\begin{center}
\includegraphics[width=17cm]{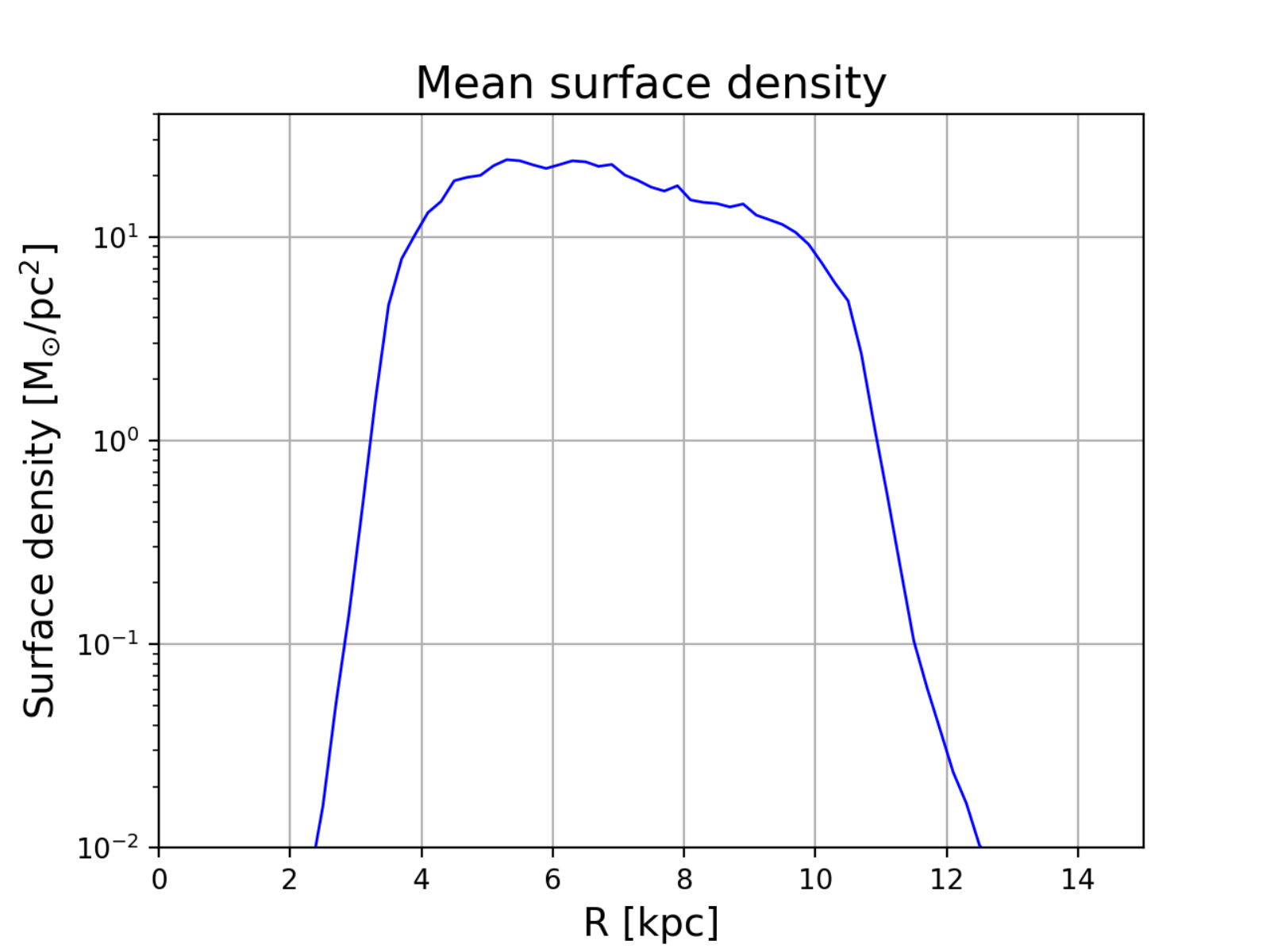}
\end{center}
\caption{
Mean surface density in \msol\ pc$^{-2}$ against radial distance in kpc of our galactic disk at the equilibrium state. The 
area that we are interested in (4 to 10 kpc) agrees well with observed Milky Way data of roughly 10 to 20 \msol\ pc$^{-2}$.
}
\label{fig:sigma}
\end{figure}

Star formation is included via a density threshold of 1000 cm$^{-3}$ and a check if the flow within the collapsing region is converging. A star formation efficiency parameter $\epsilon$ (set to 10\% in our case) determines the strength of the feedback. The total energy from stellar feedback is given by

\begin{equation}
E_{SN} = \frac{\epsilon M_{H_2}}{160 M_{\sun}}10^{51}ergs
\label{eq:esn}
\end{equation}

with each supernova contributing 10$^{51}$ erg of energy and with a supernova rate of one per 160 \msol \ of stars formed. This assumes a Salpeter initial mass function with stellar masses in the range of 0.1-100 \msol. The formation of H$_2$ is traced using the model of \cite{1996ApJ...468..269D}. For a full description of the model we refer to \cite{2011MNRAS.417.1318D}.\\

The total mass of gas is 5.78 $\times$ 10$^9$ \msol\, the number of particles is 10 $\times$ 10$^6$ and the outer disk
radius is 10 kpc. The Galaxy is modeled as a two-armed spiral, which differs from the four spiral arms of the Milky Way. However, we wanted to have prominent inter-arm regions so that we can test the impact of the cloud on the inter-arm region as well as the impact onto a spiral arm. Details of the code and test runs will be presented in a separate paper (Alig et al., in preparation).\\

The simulated galaxy reaches equilibrium after around 500 Myr. At this point in time a stable SFR of around 1.5 \msol\ yr$^{-1}$ is established, which compares well to the Milky Way \citep{2015ApJ...806...96L}.
In Fig.\ref{fig:galinit} we present the surface density distribution in \msol\ pc$^{-2}$ of the galactic disk after reaching equilibrium. Fig.\ref{fig:sigma} shows the mean surface density against radial distance for the equilibrium state at 500 Myr. The surface density within our region of interest (4 to 10 kpc) agrees well with the observational data for the Milky Way \citep{2008A&A...487..951K}. We will use this state as the starting point for our simulations of the impact
of the Smith Cloud onto the disk. By running the simulation further without including the Smith Cloud
we determine the expected SFR without impact.
The cloud parameters are adapted to the currently observed velocity and distance in $z$-direction from the Galactic Disk (which is placed within the $xy$-plane). We vary the point of impact of the cloud by shifting the initial $xy$-position of the cloud in such a way that we either hit an inter-arm region or directly the central part of a spiral arm. This is done for both cloud masses, which leads to a total of four simulations.

\section{Results}
\label{sec:num}

\begin{figure}
\begin{center}
\includegraphics[width=17cm]{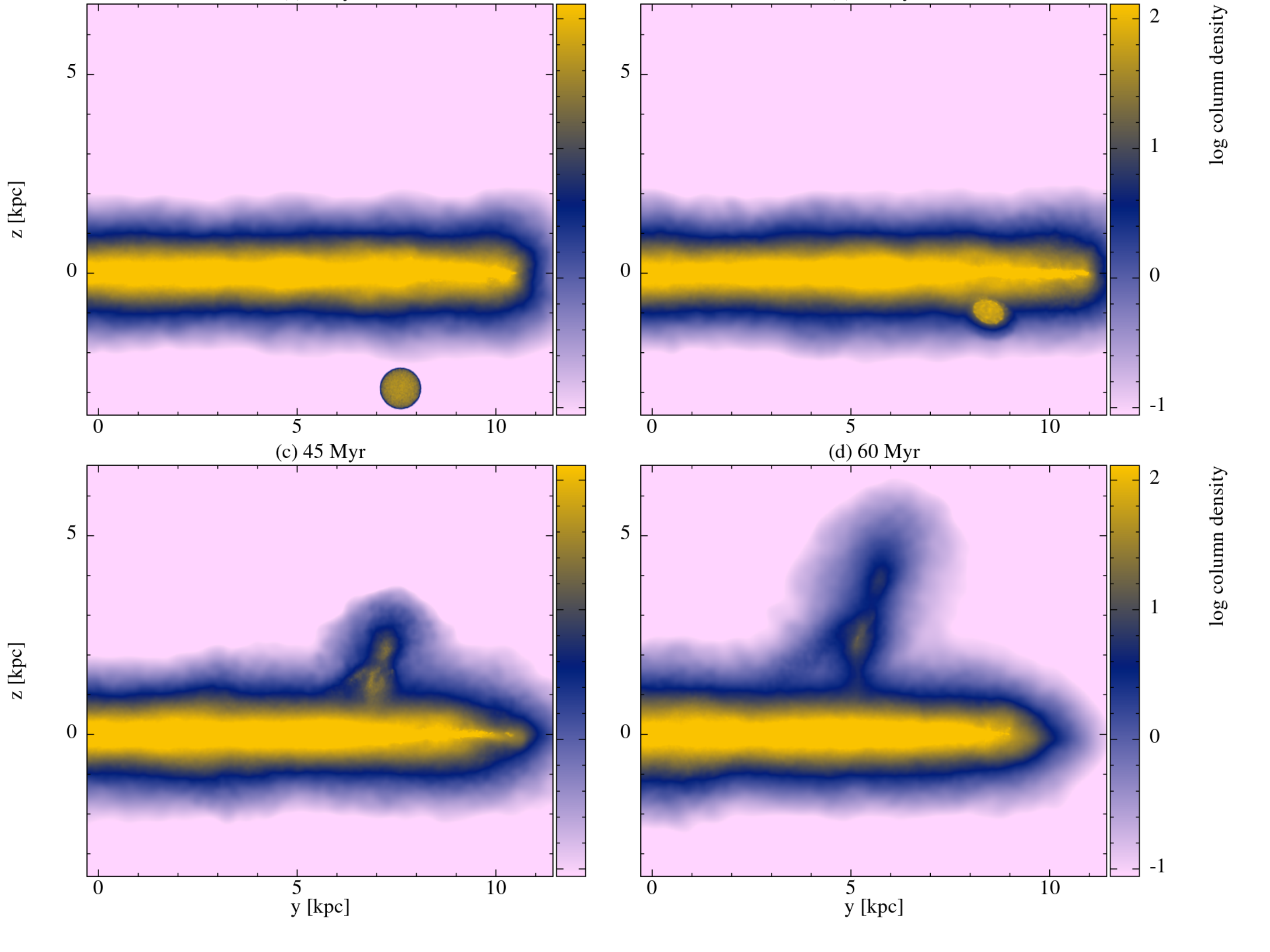}
\end{center}
\caption{
Edge-on view of the surface density of the galactic disk in in \msol\ pc$^{-2}$. The initial condition
for the 10$^7$\msol\ cloud aimed at the inter-arm region is shown in (a). In (b), at 20 Myr, the cloud touches the galactic plane. At 45 Myr shown in (c) the cloud has just passed through the plane. The top-most
high surface density spot is mostly composed of cloud gas, whereas the stream-like structure below
comprises gas dragged out of the galactic disk. In the final panel (d), at 60 Myr the cloud
becomes diffuse and a low-density stream of gas partly accretes back onto the disk and partly continues
to escape toward a higher $z$-altitude.
}
\label{fig:evol}
\end{figure}

In Fig.\ref{fig:evol} we present the evolution of the impact of the 10$^7$\ \msol\ cloud into
the inter-arm region. The plot shows the surface density in \msol\ pc$^{-2}$ in the edge-on view of the galactic disk. The initial condition in panel (a) shows the cloud at the 
observed distance from the galactic plane. In panel (b), depicting the evolution after 20 Myr, the cloud has reached the lower edge of the galactic disk and is already visibly deformed. At 45 Myr, shown in (c), parts of the cloud have broken through the disk, whilst also dragging gas of the galactic disk
outward. In the final state, shown in (d) at 60 Myr, the cloud is already strongly diffuse. 
We observe formation of a low-density stream of gas, which partly accretes back onto the galaxy and partly 
escapes to even higher $z$-altitude. The final state resembles the real shape of the Smith Cloud quite well,
implicating a previous disk passage of the Smith Cloud.
The other cases behave less extreme, with an impact of the 10$^7$\ \msol\ cloud onto the spiral arm resulting in
much less outflow and the impact of the 10$^6$\ \msol\ cloud leading to almost no gas at all at an elevated z-altitude.
The latter is almost completely absorbed during the collision.\\

\begin{figure}
\begin{center}
\includegraphics[width=17cm]{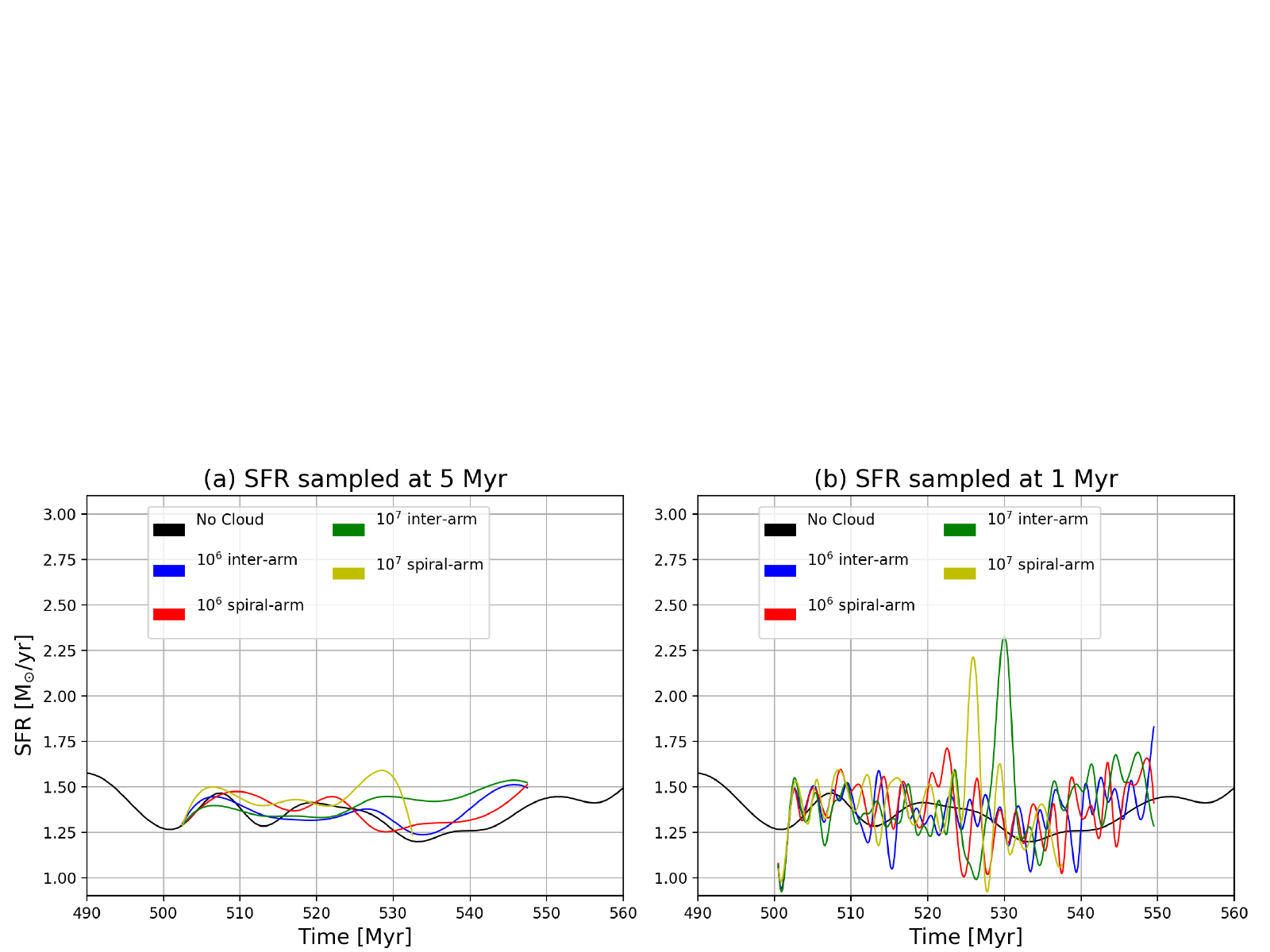}
\end{center}
\caption{
SFR in \msol\ yr$^{-1}$ over time in Myr. In the left panel (a) we show the 
SFR for the simulation without any cloud included, as well as the four cloud models, sampled
at intervals of 5 Myr. The right panel (b) shows a finer sampling rate for the simulations
including the cloud. At 5 Myr sampling the effect of the cloud impact on global SFR is largely
smoothed out. At 1 Myr sampling two short peaks emerge for the SFR of the impact of the 
high-mass cloud.
}
\label{fig:sfr}
\end{figure}

We present the SFR of all four simulations, as well as the reference-run without any cloud in Fig.\ref{fig:sfr}.
The left panel (a) shows the SFR sampled at the same interval of 5 Myr as the long term simulation 
without an external cloud. Panel (b) shows the SFR for the simulations including the cloud sampled at 
the interval of 1 Myr. In both cases the simulations performed with the 10$^6$\ \msol\ low-mass cloud
show nearly no deviation from the run without any cloud included. In the case of the 10$^7$\ \msol\ cloud 
the 5 Myr sampled SFR also shows no large deviation beyond the natural fluctuation of the SFR for the run without any cloud included. However, when sampled at 1 Myr two peaks emerge with an increase in SFR of almost
1\ \msol\ yr$^{-1}$. The difference in the SFR for the inter-arm impact simulation and the 
spiral-arm impact simulation is negligible. The only visible difference is the time delay in
the star formation peak of the 10$^7$\ \msol\ case
due to the cloud encountering a large amount of gas earlier in the spiral-arm region, whereas the cloud has to penetrate deeper into the disk in the inter-arm region in order to trigger a local starburst.\\

The area of impact of the Smith Cloud is small compared to the whole area of the disk. 
However, if this area accounts for almost double the normal Milky Way SFR for some time, the resulting starburst should be observable. Still, there is a strong dependence on the cloud mass, and we made assumptions for our cloud model that should at least favor an increased SFR. Thus, in reality we would expect the effect on global SFR to be smaller even in the high-mass case.\\

\begin{figure}
\begin{center}
\includegraphics[width=17cm]{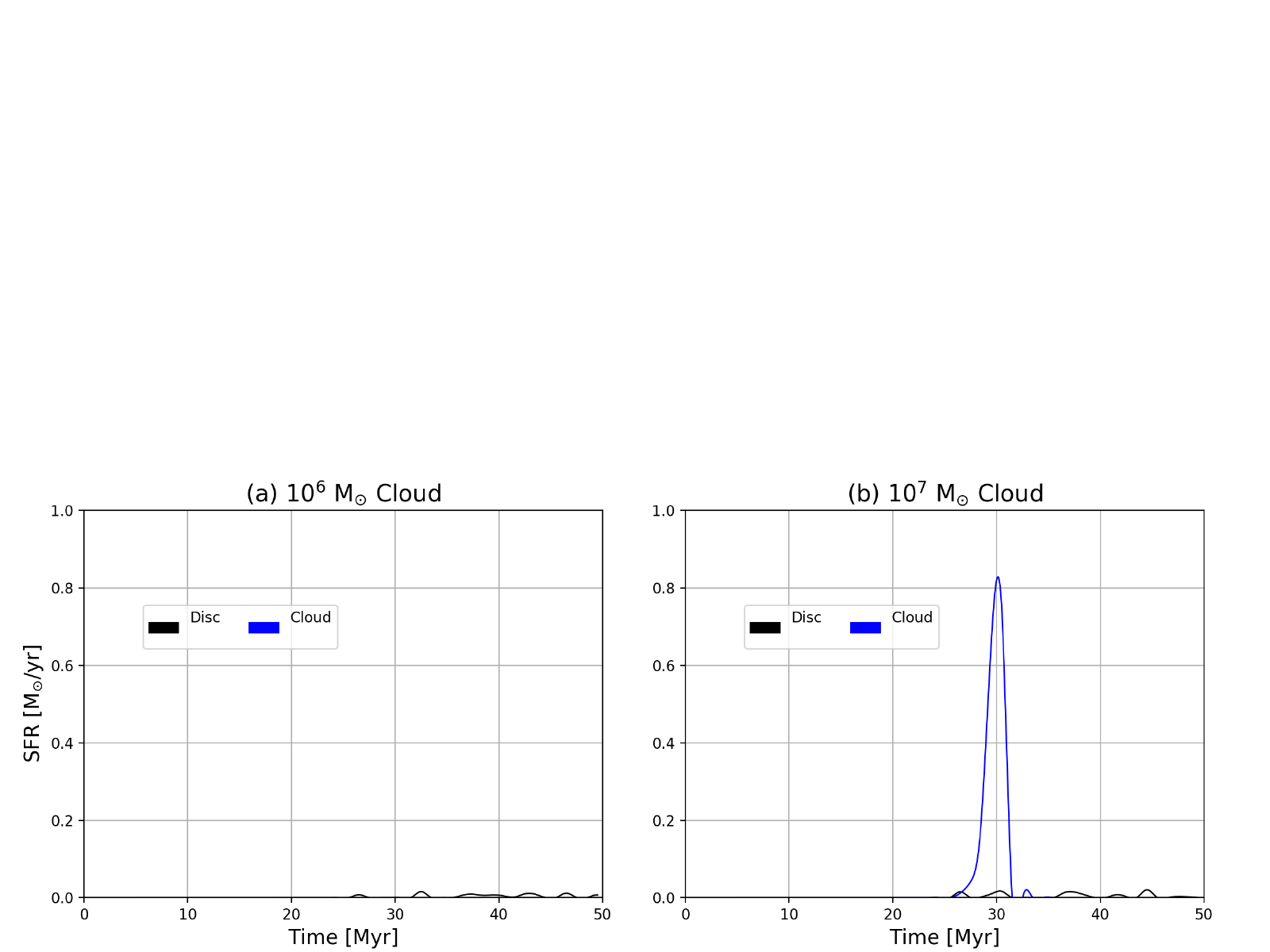}
\end{center}
\caption{
Local SFR in \msol\ yr$^{-1}$ over time in Myr for the galactic disk gas within the cloud impact area and 
for the cloud gas only. Panel (a) shows the inter-arm hit for the low-mass cloud and panel (b) the inter-arm
hit for the high-mass cloud.
During the time span of 25 to 35 Myr the cloud reaches the galaxy mid-plane and crosses the patch of gas that we follow.
The result for the high-mass cloud impact (b) shows a clear signal within the cloud, however the disk gas seems to be
largely unaffected. By comparison, the 
result for the low-mass cloud (a) shows no increase in the SFR of the cloud or the disk during the impact.
}
\label{fig:sfrloc}
\end{figure}

Finally, we look at the local SFR of the cloud impact zone. Even if the global SFR is not 
affected by the cloud impact, the impact area of the cloud inside the galactic disk could show an increased SFR. We compare the local SFR for the 10$^6$\ \msol\ and the 10$^7$\ \msol\ case in Fig.\ref{fig:sfrloc}. The plot shows the SFR in \msol\ yr$^{-1}$ over time in Myr for the ISM patch, which is positioned at the disk mid-plane when the cloud crosses the mid-plane. In addition,
we also show the SFR for the cloud gas only.
We marked the disk gas particles of the local ISM patch close to the cloud mid-plane crossing point in time, thus the SFR will be zero for the time before because all of the gas that has already been consumed by stars before is no longer included.
The impact of the $10^7$\ \msol\ cloud can be clearly seen when it starts entering the patch at around 25 Myr. 
However, only the cloud gas seems to be forming stars, therefore the ISM itself is largely unaffected.
The low-mass cloud, however, shows no
fluctuations beyond the background level. Thus we would not expect a 
starburst-like event in this case, even specifically at the area of entry of the cloud into the Milky Way disk.

\section{Summary}
\label{sec:sum}

We have simulated the impact of the Smith Cloud onto the gaseous galactic disk. The Smith Cloud is modeled by a sphere placed at the tip position of the original cloud, which contains the mass of 
the whole extended cloud. This assumption should at least favor an increased SFR peak (not necessarily the total amount of star formation), because more mass can impact at once compared to an elongated cloud which is incorporated into the disk and transported away from the impact point (thus decoupling it from the influence of the gas falling in later) over a larger timescale. As the cloud mass is not known with absolute certainty, we assume two masses for the cloud: a lower
limit of 10$^6$\ \msol\ and an upper limit of 10$^7$\ \msol. To test the influence on galactic SFR, 
we simulated the collision of the cloud onto a inter-arm region and the collision onto a spiral-arm
region. Our main findings are as follows.\\

\begin{itemize}
 \item The low-mass cloud simulations produce no visible increase in the global or even 
 local SFR of the galaxy. The cloud gas is nearly completely absorbed into the disk and 
 no gas can escape toward the opposite side of the impact.\\
 
 \item The high-mass cloud simulations show a short-term increase of the SFR. This can amount to up to
 double the regular SFR of the galaxy. This increase is confined to the small
 impact region, so the effect should be observable as a localized starburst.\\
 
 \item Our assumptions favor an increased SFR peak. The actual cloud is rather extended and  contains a lower amount of mass compared to our high-mass model. Thus in reality the local SFR peak should be lower but more extended over time. However, we also assumed the cloud to be homogenous, when in reality the cloud could contain
 local patches of high-density gas within the tip. Those patches could be dense enough for star-formation to occur when the cloud gets compressed while entering the gaseous disk. Simulating a cloud with sub-structure is currently beyond the resolution limit of our galaxy-scale simulationsm but should be considered in future work.
 
\end{itemize}

\section*{Acknowledgments}
\label{sec:ack}

We would like to thank the anonymous referee for suggestions and hints that helped to substantially improve this paper.
Computer resources for this project have been provided by the Gauss Center for Supercomputing/Leibniz Supercomputing Center under grant: di25mim. The surface density plots have been created using the publicly available SPH visualization tool SPLASH by D.J. Price \citep{2007PASA...24..159P}.

\bibliographystyle{aasjournal}

\label{lastpage}
\end{document}